\documentclass[a4paper,fleqn,usenatbib]{mnras}

\usepackage{newtxtext,newtxmath}
\usepackage[T1]{fontenc}
\usepackage{ae,aecompl}

\usepackage{graphicx}	% Including figure files
\usepackage{amsmath}	% Advanced maths commands
\usepackage{amssymb}	% Extra maths symbols
\usepackage[]{algorithm2e}
\usepackage[caption=false]{subfig}
\graphicspath{{figures/}}

\usepackage{etoolbox}
\makeatletter
\patchcmd\@combinedblfloats{\box\@outputbox}{\unvbox\@outputbox}{}{%
	\errmessage{\noexpand\@combinedblfloats could not be patched}%
}%
\makeatother

\title[Spatial dispersion of light rays]{Spatial dispersion of light rays propagating through a plasma in Kerr spacetime}

\author[T. Kimpson et al.]{
Tom Kimpson,$^{1}$\thanks{E-mail: tom.kimpson.16@ucl.ac.uk}
Kinwah Wu ,$^{1}$
and Silvia Zane $^{1}$
\\
$^{1}$ Mullard Space Science Laboratory, University College London. Holmbury St. Mary, Dorking, Surrey, RH5 6NT, UK
}

\date{Accepted XXX. Received YYY; in original form ZZZ}

\pubyear{2018}

\begin{document}
\label{firstpage}
\pagerange{\pageref{firstpage}--\pageref{lastpage}}
\maketitle

\begin{abstract}
We investigate the propagation of light through a plasma on a background Kerr spacetime via a Hamiltonian formulation. The behaviour of light when propagating through a vacuum and through a plasma is not the same; the convolution of gravitational and plasma effects gives rise to a dispersion in both space and time. The magnitude of the dispersion is a strong function of both the ray frequency and impact parameter. We discuss implications for the detection of gravitationally bent pulsar beams near the Galactic centre.
\end{abstract}

\begin{keywords}
gravitation -- pulsars -- black hole physics
\end{keywords}

%%%%%%%%%%%%%%%%%%%%%%%%%%%%%%%%%%%%%%%%%%%%%%%%%%

%%%%%%%%%%%%%%%%% BODY OF PAPER %%%%%%%%%%%%%%%%%%

\section{Introduction}
\label{sec:level1}

Black holes (BHs) are mathematical solutions to Einstein's field equations. The most simple solution of an uncharged, non-spinning BH - the Schwarzschild solution \citep{Schwarzschild1999}  - was discovered soon after Einstein's theory of general relativity (GR) was proposed, whilst the solution for the spinning BH - the Kerr solution \citep{Kerr1963} - was discovered much later. Originally BH's were conceived as purely theoretical constructs, rather than astronomical bodies. However, over time, astronomers began to become convinced as to the existence of astrophysical BHs: in X-ray binaries \citep{Bolton1972,Webster1972,Generozov2018, Hailey2018}, at the centre of globular clusters \citep{Ferrarese2000, Gebhardt2002, Wrobel2018}, and in the hearts of massive galaxies as the engines of Active Galactic Nuclei \citep{Salpeter1964, Zeldovich1964,Madejski2002}. Indeed, our own galaxy is thought to host a supermassive black hole (BH) of mass $4.3 \times 10^{6} M_{\odot}$ at the Galactic centre \citep{Gillessen2009}. The most recent evidence for BH's as physical objects was provided by multiple gravitational wave observations of merging black holes by the LIGO/VIRGO experiment \citep{LIGOO2018}.

Astrophysical BHs provide superb environments for tests of fundamental physics and astrophysics. Whilst GR has been enormously successful historically \citep[see discussion in][]{Will2014}, a range of open questions do persist. Astronomical observations of BHs can be used for investigating alternative theories of gravity \citep[e.g. scalar-tensor theories][]{Esposito2009, Liu2014}, probing deviations from the Kerr solution \citep[`bumpy black holes'][]{Yagi2016} and even testing quantum gravity theories \citep{Estes2017}. From an astrophysical perspective, central supermassive BHs appear to play a key role in galaxy formation and evolution \citep{Cattaneo2009}. Moreover the observed relation between the nuclear BH mass and the galactic stellar velocity dispersion  \citep[the $M-\sigma$ relation,][]{Ferrarese2000} suggests that BHs and galaxies coevolve \citep{Lamastra2010,Schawinski2012, Heckman2014}, and hints to the existence of intermediate mass BHs \citep{Koliopanos2017, Mezcua2017}. BHs are also postulated to contribute to the cosmic X-ray background \citep{Gilli2013}, and power jets in X-ray binaries \citep{Fender2004,Fender2009}. The advent of gravitational wave and multimessenger astronomy will further enhance the scientific return from observations of BH systems.

The BH solutions by Schwarzschild and Kerr represent vacuum solutions. However, these astrophysical BHs do not exist in a vacuum but are instead surrounded by a plasma distribution \citep{Psaltis2012, Eatough2013}. It is therefore of astrophysical interest to determine the impact of this plasma on any astronomical observations. Photons that are emitted close to the BH and propagate through a plasma to be received by an observer suffer from significant transfer effects. Ignoring the polarisation of the light, we may classify the transfer effects into two general categories: (i) those of relativistic and/or gravitational nature \citep[see e.g.][]{Fuerst2004,Saxton2016}, and (ii) those due to the interaction between the emission and the line-of-sight material \citep[cf. the relativistic fluids near a black holes see e.g.][]{Fuerst2007,Younsi2012}. If only one category is significant, 
the transfer effects can be accounted for in a straightforward manner, at least in principle. For the former, a covariant ray-tracing formulation will be adequate to determine the convolution of the effects due to gravitational lensing, time dilation, velocity induced intensity boost and relativistic Doppler frequency shift, provided that the motion of the emitter is specified. For the latter, the frequency dependent dispersion of the emission can be calculated, if the density distribution of the plasma along the line-of-sight is known. When both of them are present, we cannot simply ``add'' their effects together or use a simple convolution of the effects assuming that the geodesics (along which the rays are traced) determined in the vacuo condition are still applicable. The deflection/bending of the ray is now frequency dependent, a consequence of the interaction with the plasma as the ray propagates in a curved spacetime.  \newline 

\noindent The influence of plasma on photon rays has been investigated by a number of authors around both spinning (Kerr) and non-spinning (Schwarzschild) black holes, particularly with application towards the impact on the black hole shadow or the photon sphere \citep{Bisnovatyi-Kogan2010, Atamurotov2015, Liu2016, Liu2017, Perlick2017,Benavides2018, Dokuchaev2018, Huang2018}. In this work we demonstrate the phenomenon of spatial dispersion of the ray that is introduced as a consequence of the plasma. We show that this dispersion is significant at typical radio frequencies for rays which pass sufficiently close to the central BH and consider the impact on the time delay of the ray due to the convolution between the delay induced by the plasma and the delay due to the changing ray path. It is expected that at the Galactic centre there should exist a large population of pulsars, numbering up $\sim 10^4$ within 1 pc of Sgr A* \citep{Macquart2015,Rajwade2017}. The detection of such systems is highly desirable for the purposes of testing key questions of GR with high precision in an extreme parameter space \citep{Liu2012, Psaltis2016}.  However, currently no pulsars have been detected within 1 pc of Sgr A* \citep[see e.g. ][]{Chennamangalam2014, Dexter2014}. We discuss potential astrophysical implications of the dispersion induced by the plasma, particularly with regard towards the detection and timing of strongly gravitationally bent rays from Galactic centre pulsars. 

The structure of the paper is as follows. In Section \ref{sec:light_propagation} we outline the equations of motion for a photon propagating through a plasma in Kerr spacetime and discuss necessary conditions on the plasma density distribution to ensure their integrability. In Section \ref{sec:dispersions} we solve these equations for a specific plasma density distribution and show that the presence of plasma induces both a spatial and temporal dispersion in gravitationally bent rays at typical radio frequencies. Discussion regarding the detection of strongly bent rays from Galactic centre pulsars is made along with concluding remarks in Section \ref{sec:discussion}.

\section{Propagation of light rays in a Kerr spacetime} 
\label{sec:light_propagation} 

We adopt the natural units, with $ c=G=\hbar = 1$, and a $(-,+,+,+)$ metric signature.  
Unless otherwise stated, a c.g.s. gaussian unit system is used 
in the expressions for electromagnetic properties of matter. 
The gravitational radius of the black hole $r_{\rm g} = M$ 
and the corresponding Schwarzschild radius $r_{\rm s} = 2M$, where $M$ is the black-hole mass. We will adopt a normalization that the black-hole mass is unity.  
A comma denotes partial derivative (e.g.$\;\! x_{,r}$), 
and a semicolon denotes covariant derivative (e.g.$\;\! x_{;r}$). 

%%%%%%%%%%%%%%%%%%%%%%%%%%%%%%%%%%%%%%%%%%%%%%%%%%%%
\subsection{Light propagation under gravity}
The vacuo spacetime around a rotating astrophysical black hole - neglecting any contribution from stellar objects or the surrounding medium - is described by the Kerr metric. In Boyer-Lindquist coordinates, its spacetime interval is given by 
\begin{align}
	{\rm d}s^2 &= -\left(1 - \frac{2r}{\Sigma}\right) {\rm d}t^2 
	- \frac{4ar \sin^2 \theta}{\Sigma}\ {\rm d}t \;\! {\rm d}\phi 
	+ \frac{\Sigma}{\Delta}{\rm d}r^2  \nonumber \\ 
	\hspace*{1.2cm} &+ \Sigma\ {\rm d} \theta^2 + \frac{\sin^2 \theta}{\Sigma} \left[(r^2+a^2)^2 - \Delta a^2 \sin^2 \theta \right] {\rm d}\phi^2 \ , 
	\label{eq:kerr_metric} 
\end{align}
where $\Sigma = r^2 +a^2 \cos^2 \theta$, $\Delta = r^2 - 2r +a^2$, and $a$ is the black-hole spin parameter. The Kerr spacetime possesses two Killing vectors $\xi^{t}, \xi^{\phi}$, related to temporal and axial diffeomorphisms of the vacuum metric $g^{\mu \nu}$. A general Killing vector $\xi^{\mu}$ satisfies Killing's equation $\xi^{(\mu;\nu)} = 0$. The inner product of a Killing tensor $\xi^{\mu}$ with a tangent vector $p_{\mu}$ is conserved along a geodesic, i.e. if $K =\xi^{\mu}p_{\mu} $ then $\dot{K} = 0$, where ``$\cdot$'' denotes differentiation with respect to an affine parameter. The two Killing vectors are associated with the conservation of energy $E$ and angular momentum $L_z$ (specifically the projection of the particle angular momentum along the black hole spin axis) as,
\begin{eqnarray}
	E = - \xi^{t} p_t \ ,
\end{eqnarray}
\begin{eqnarray}
	L_z =  \xi^{\phi} p_{\phi} \ .
\end{eqnarray}
In addition, the Kerr spacetime admits a rank-2 Killing tensor 
\begin{eqnarray}
	K^{\mu \nu} = 2 \Sigma l^{(\mu} n^{\nu)} + r^2 g^{\mu \nu} \ ,
\end{eqnarray}
where $l^{\mu}, n^{\nu}$ are the principal null vectors. This Killing tensor obeys the Killing tensor equation,
\begin{eqnarray}
	\nabla^{(\gamma }K^{\mu \nu)} = 0 \ .
\end{eqnarray}
The Killing tensor is related to a conserved quantity which is not associated with a spacetime symmetry - the Carter Constant ($Q$) - and was originally derived from the separability of the Hamiltonian in $r$ and $\theta$ terms \citep{Carter}. The exact physical meaning of $Q$ is discussed in \citet{DeFelice1999, Rosquist2009}. The rest mass of the particle (i.e. $H=0$ for photons) is also conserved. Determining the motion of a photon in Kerr spacetime therefore reduces to a problem of quadratures whereby we have four ordinary differential equations (ODEs) for each of the spacetime coordinate variables $(\dot{t}, \dot{r}, \dot{\theta}, \dot{\phi})$,  
and four associated constants of motion $E,$ $L_z$, $Q$, and $H$. The system of equations is then integrable. \newline 

\noindent For photon propagation in vacuo, the Hamiltonian is,
\begin{eqnarray}
	H(x^{\mu},p_{\nu}) = \frac{1}{2} g_{\mu \nu} p^{\mu} p^{\nu } = 0   \ , 
\end{eqnarray} 
where $x^{\mu}$ are the coordinate variables and $p_{\nu}$ the conjugate 4-momenta. The corresponding equations of motion, Hamilton's equations, are then
\begin{eqnarray}
	\dot{x^{\mu}} = \frac{\partial H}{\partial p_{\mu}} \, , \hspace*{0.2cm} \dot{p}_{\mu} = -\frac{\partial H}{\partial x^{\mu}} \ .
\end{eqnarray} 
For photon propagation through a cold, non-magnetized electron-proton plasma, 
the Hamiltonian has an additional term proportional to the electron plasma frequency $\omega_{\rm{p}}(x^\mu)$:  
\begin{eqnarray}
	H(x^\mu,p^\mu) = \frac{1}{2} \left[g_{\mu \nu} p^{\mu} p^{\nu } + \omega^2_{\rm{p}}(x^\nu)\right] =0 \ ,
\end{eqnarray}  
in the geometrical optics approximation \citep{Synge1960}. Thus, in the presence of a plasma the preceding argument regarding integrability needs modification.

%%%%%%%%%%%%%%%%%%%%%%%%%%%%%%%%%%%%%%%%%%%%%%%%%%%%
\subsection{Propagation of light in a cold plasma with azimuthal symmetry}

If the plasma frequency is independent of $t$ and $\phi$,  
i.e. $\omega_{\rm p} = \omega_{\rm p} (r, \theta)$,  
the Hamiltonian is still stationary and axisymmetric.  
Although the photon rest mass remains null, the inclusion of the plasma-frequency term breaks the separability of the Hamiltonian, 
implying that the Carter constant for deriving the geodesics in a vacuum spacetime
is no longer properly defined. 
To illustrate, we consider the following: 
\begin{align}
	&-\frac{E^2}{\Delta} \left[(r^2+a^2)^2 - \Delta a^2 \sin^2 \theta \right]
	+ \frac{4 a r EL_z}{\Delta} \nonumber \\    
	\hspace*{0.75cm} &+ p_r^2 \Delta +p_{\theta}^2 + \frac{L_z^2}{\sin^2 \theta} \left(1 - \frac{a^2 \sin^2 \theta}{\Delta}\right) \nonumber \\  
	&+ r^2 \omega_{\rm{p}}^2(r,\theta) +a^2\cos^2\theta\, \omega_{\rm{p}}^2(r,\theta)= 0 \ , 
\end{align} 
which adopts the definitions of $E$ and $L_z$ above and the expression for $H$.  
In the absence of a plasma, $\omega_{\rm{p}} = 0$, 
and this expression is separable into $r$ and $\theta$ terms, 
which can then be used to define the Carter constant. 
For general plasma density distributions, where the plasma frequency, $\omega_{\rm{p}}(r,\theta)$, has a spatial dependence,  
the equation is not separable in terms of the usual co-ordinate variables. Moreover, the Carter constant as derived in the vacuum situation is no longer a constant along the geodesic \citep[see e.g.][]{Perlick2017}.

The underlying physics can be elucidated by comparing a plasmic-Hamiltonian,  
\begin{eqnarray}
	g_{\mu \nu} p^{\mu} p^{\nu} + \omega_{\rm{p}}^2 = 0  \ , 
\end{eqnarray}
with the Hamiltonian for a particle of mass $m$ travelling through a vacuum,
\begin{eqnarray}
	g_{\mu \nu} p^{\mu} p^{\nu} + m^2 = 0  \  . 
\end{eqnarray}
The plasma frequency effectively plays the role of a particle mass in a vacuum spacetime. For a uniform plasma distribution, 
$\omega_{\rm p}$ is a constant and hence the Hamiltonian is separable. Otherwise, the photon essentially acquires an effective mass which varies during its propagation in the plasma. 
Consequently, we have three constants of motion for four differential equations. 
The equations of motion can be integrated  
only when the remaining constant of motion (or symmetry) is identified. \newline

\noindent The plasma frequency of a cold non-magnetised plasma is
\begin{eqnarray}
	\omega^2_{\rm{p}} = C\;\! n \ ,
\end{eqnarray}
where $C = 4\pi e^2/m_e$ with $m_e$ the electron mass,  $e$ the electron charge, and $n$ is the electron number density. We notice that for stationary axisymmetric plasma distributions such that 
\begin{eqnarray}
	\omega^2_{\rm{p}} = \beta \ \frac{f(r)+g(\theta)}{\Sigma}  \ ,
	\label{eq:plasma}
\end{eqnarray} 
the Hamiltonian is separable \citep{Perlick2017}. This distribution corresponds to plasma density distributions 
with independent radial and polar dependence through the additive contribution by the two terms $f(r)$ and $g(\theta)$.
Although this particular form restricts the description of general density distributions, it retains certain desirable properties from the perspective of astrophysical modelling, as with an appropriate choices of $f(r)$ and $g(\theta)$ we may describe the key features of an axisymmetric plasma, such as inverse radial dependence, dominance of radial terms at large radii, maximal value in equatorial plane, etc. Going forward, we adopt this particular functional form of  $\omega_{\rm{p}}$ in our demonstrative calculations.

%%%%%%%%%%%%%%%%%%%%%%%%%%%%%%%%%%%%%%%%%%%%%%%%%%%%
\subsection{Frequency dependent ray-tracing}
It follows that the complete set of equations of motion is given, via Hamilton's equations, as 
\begin{align}
	\dot{t} &= E + \frac{2r(r^2 +a^2)E - 2arL_z}{\Sigma \Delta}  \  ;  \\ 
	\dot{r} &= \frac{p_r \Delta}{\Sigma}  \ ;  \\ 
	\dot{\theta} &= \frac{p_{\theta}}{\Sigma} \ ;    \\ 
	\dot{\phi} &= \frac{2arE + (\Sigma - 2r)L_z\csc^2\theta}{\Sigma \Delta} \ ;   \\ 
	\dot{p}_{\theta} &= \frac{1}{2 \Sigma} \left[-C g (\theta)_{, \theta} -2a^2 E^2 \sin \theta \cos \theta + 2L_z^2 \cot \theta \csc^2 \theta \right] \ ; \\ 
	\dot{p}_r &= \frac{1}{\Sigma \Delta} \big[-\kappa(r-1) +2r(r^2+a^2)E^2 - 2aEL_z \\ 
	&-  \frac{C f(r)_{,r} \Delta}{2}  -C  (r-1) f(r)\big]\\ 
	&-\frac{2 p_r^2(r-1)}{\Sigma}  \ ;
\end{align}
where $\kappa = p_{\theta}^2 +  E^2a^2 \sin^2 \theta + L_z^2 \csc ^2 \theta + a^2 \omega_{\rm{p}}^2 \cos^2 \theta$. The differential equations are integrated `backwards-in-time' from the observer image plane to the black hole, using a fifth-order Runge-Kutta-Fehlberg algorithm with adaptive step-size \citep{Press1992}. The centre of the observer's image plane is defined at some location $(r_{\rm{obs}}, \theta_{\rm{obs}}, \phi_{\rm obs})$ where $r_{\rm{obs}}$ is the distance from the black hole centre and $\theta_{\rm obs}$ the angle from the positive black hole $z$-axis. Since the Kerr metric is axisymmetric we can set $\phi_{\rm obs}=0$ (see Fig 3.4 of \cite{YounsiThesis} for an illustration of the coordinate system used). The observer distance $r_{ \rm obs}$ is chosen so as to be sufficiently large such that the observer's grid can be considered as a Euclidean grid with zero curvature, and all rays are perpendicularly incident on this grid. Throughout this work we set our observer grid at $r_{\rm obs} = 10^4\;\! r_{\rm g}$. We can approximate the deviation from Minkowski spacetime via the Kretschman scalar, which for a Kerr spacetime is \citep{Henry2000},
\begin{eqnarray}
	\mathcal{K} = \frac{48}{\Sigma^6}(r^6 - 15a^2r^4 \cos^2 \theta + 15 a^4 r^2\cos^4 \theta - a^6 \cos^6) \ .
\end{eqnarray} 
In the limit of large $r \, (=r_{\rm obs})$,
\begin{eqnarray}
	\mathcal{K} \sim \frac{48}{r^6} \sim 10^{-24} \ ,
\end{eqnarray} 
which is much smaller than typical numerical precision and so we are well-justified as taking the observer plane as Euclidean.
The two integration constants, $E$ (energy at infinity) and $L_z$ (the azimuthal component of the angular momentum), 
can be determined by the initial conditions using the following relations:  
\begin{align}
	E^2 &= (\Sigma - 2r) \left(\frac{\dot{r}^2}{\Delta} + \dot{\theta}^2 
	+ \frac{ \omega_{\rm{p}}^2}{\Sigma}\right) + \Delta \dot{\phi}^2 \sin^2 \theta   \ ;   \\ 
	L_z &= \frac{(\Sigma \Delta \dot{\phi}-2arE)\sin^2\theta}{\Sigma - 2r}   \ . 
\end{align}  
An impact parameters quantifies the perpendicular distance between the centre of the black hole and the asymptote of the tangent line to the ray that converges at the observer. We refer to the impact parameters $\alpha, \beta$ to describe this distance in the $y$ and $z$ directions respectively. The complete specification of the initial conditions is described in the Appendix. 
\section{Dispersions induced by plasma}
\label{sec:dispersions}
In order to solve the equations numerically and assess the degree of dispersion, it is necessary to choose a model for the plasma frequency $\omega_{\rm p}$, which is equivalent to choosing an electron number density distribution for the plasma. We approximate a real astrophysical plasma around a black hole using the semi-analytical model of \cite{Broderick2005} where the parameters were determined via simultaneous fitting of spectral and polarization data from Sgr A*. This description is also used for Galactic centre studies in \cite{Psaltis2012}. Within this model the electron density profile is given by,
\begin{eqnarray}
	n = n_0 r^{-1.1} \ .
	\label{eq:density}
\end{eqnarray}
This is equivalent to setting $f(r) = r^{0.9}$ and $g({\theta}) = 0$, such that,
\begin{eqnarray}
	\omega_{\rm p}^2 = \frac{Cr^{0.9}}{\Sigma} \ .
\end{eqnarray}
To ensure integrability it is necessary to contain the $\theta$ dependence as implicit in $\Sigma$, but we take this form as a  decent first-order approximation to the Galactic centre plasma. In the equatorial and zero spin case, we recover the form of Eq. \ref{eq:density}. The best fit parameters to the model are $n_0 = 3.5 \times 10^7$ cm$^{-3}$ \citep{Broderick2011}, and similar normalizations are commonly used in the literature \citep[e.g.][]{Psaltis2012, 2009ApJ...706..497M}. In relativity, to associate a frequency to a photon requires the specification of the observer's 4-velocity $u^{\mu}$. The frequency as measured by an observer is given by the frame-invariant quantity, $\nu = p_{\mu} u^{\mu}$. There also exists a frequency associated with the temporal Killing vector $\nu_{K} = p_t \xi^{t}$ which is the frequency of a stationary observer at infinity, and is conserved along the ray path. 
\subsection{Spatial Dispersion}
\begin{figure*} 
	\subfloat[$\alpha = -9$]{\includegraphics[width=0.333\textwidth]{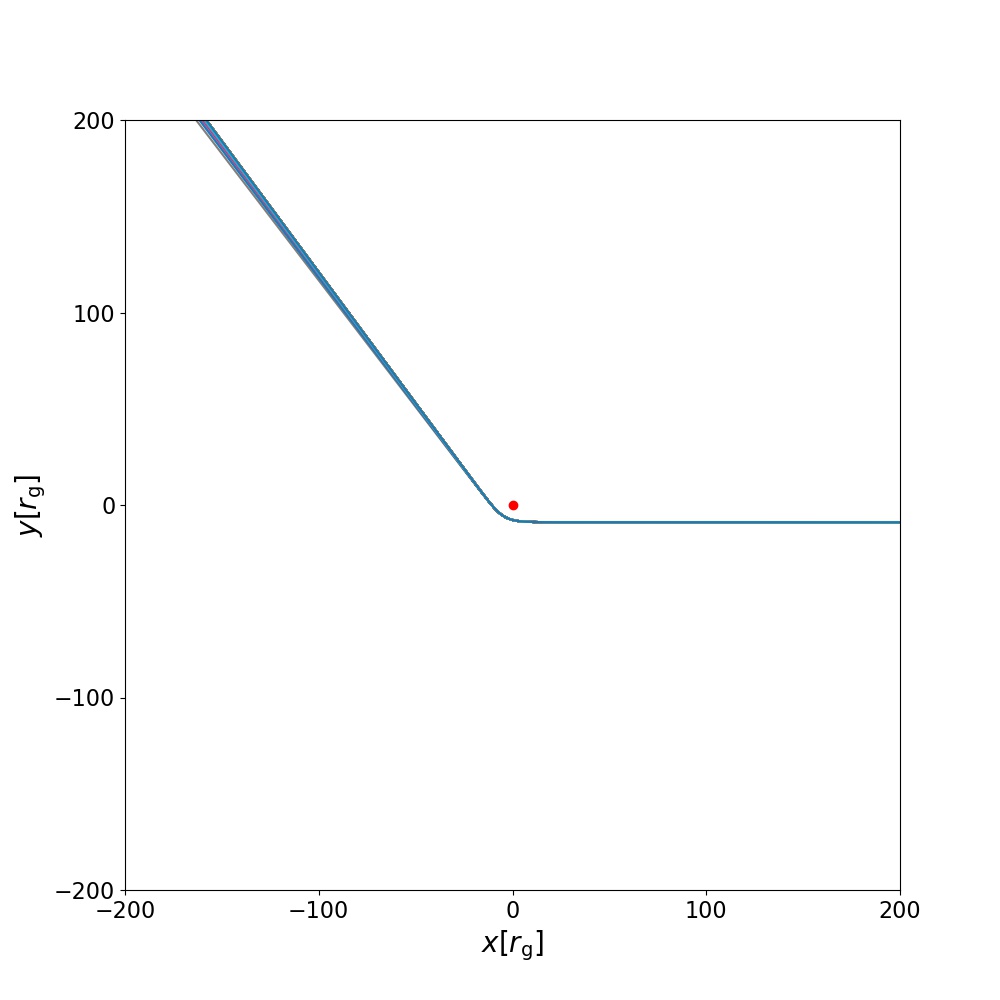}}
	\subfloat[$\alpha = -8$]{\includegraphics[width=0.333\textwidth]{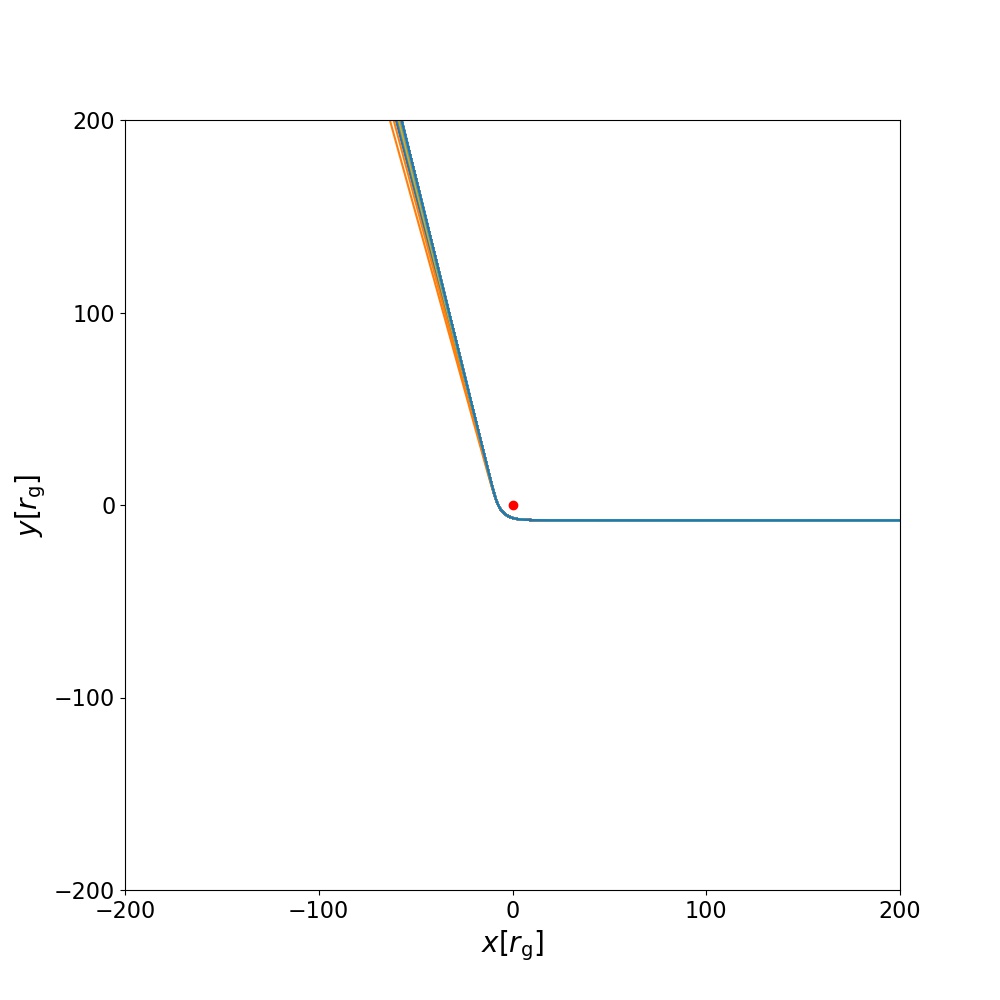}}
	\subfloat[$\alpha = -7$]{\includegraphics[width=0.333\textwidth]{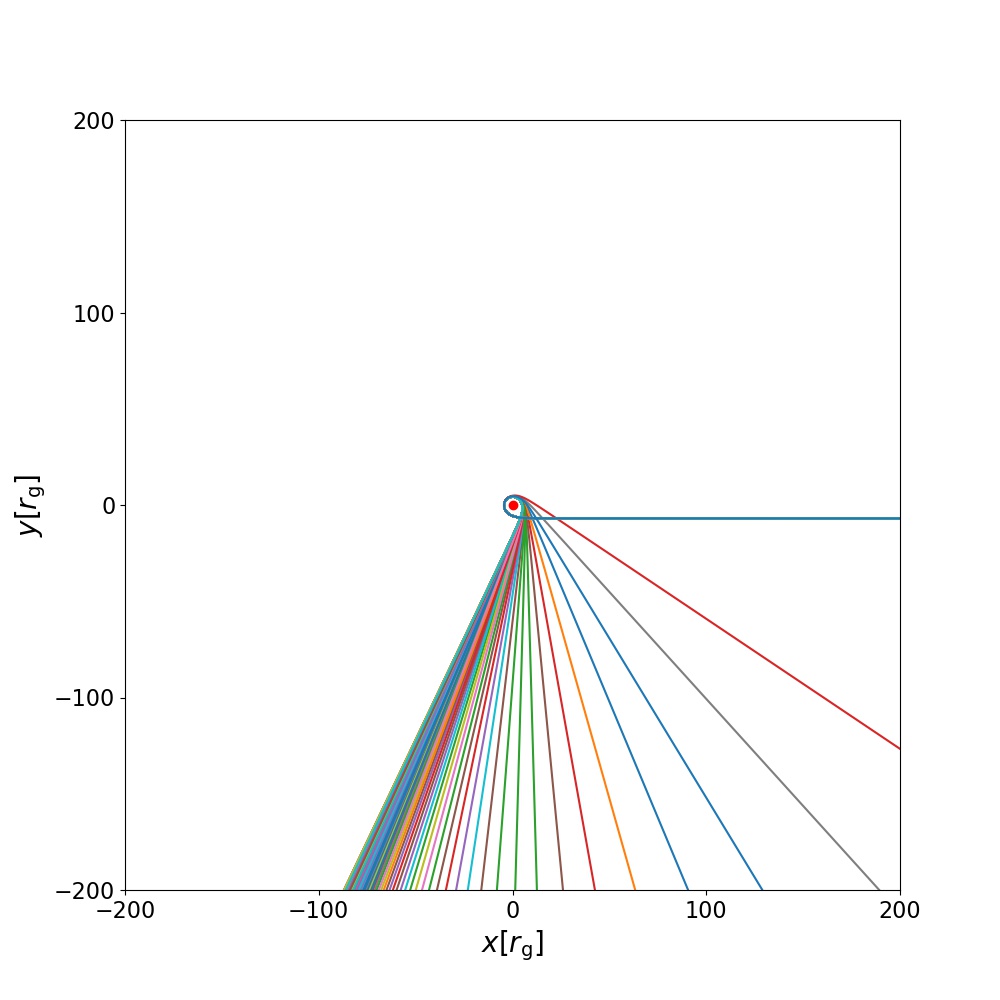}}\\
	\subfloat[$\alpha = +5$]{\includegraphics[width=0.333\textwidth]{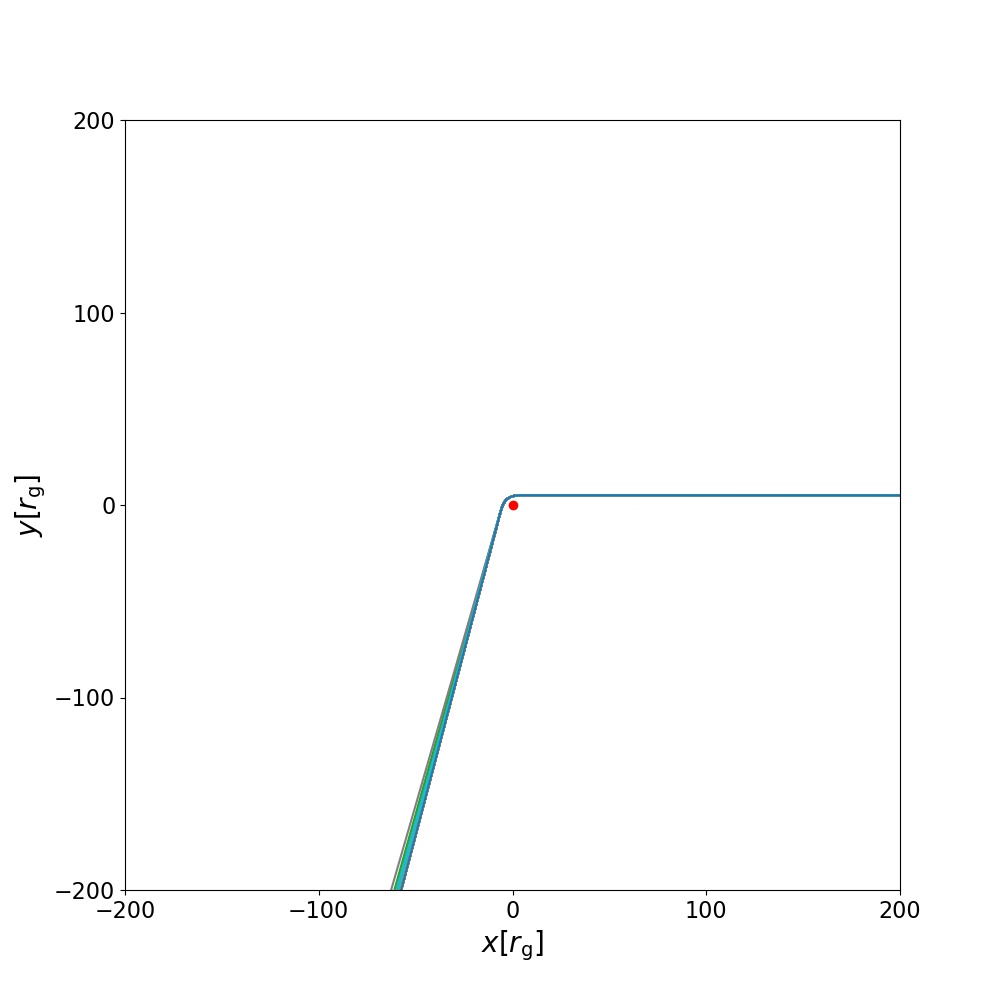}}
	\subfloat[$\alpha = +4$]{\includegraphics[width=0.333\textwidth]{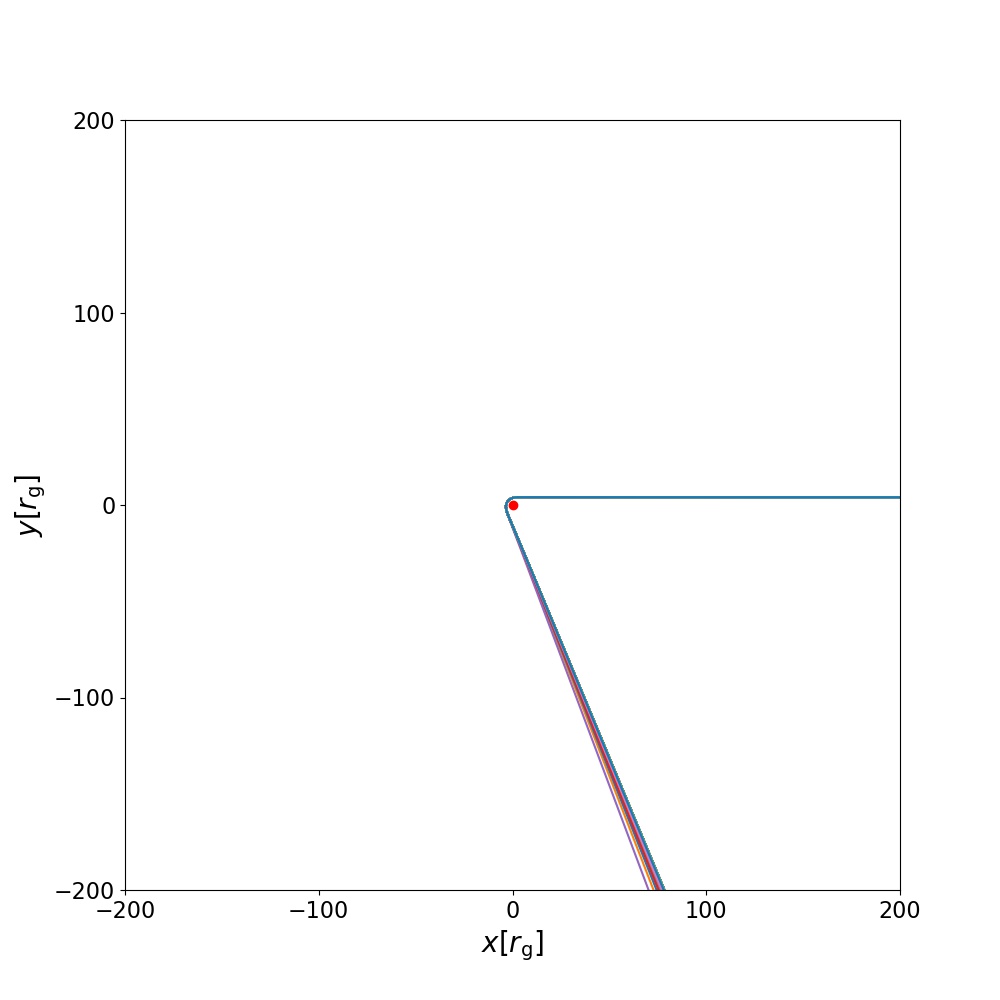}}
	\subfloat[$\alpha = +3$]{\includegraphics[width=0.333\textwidth]{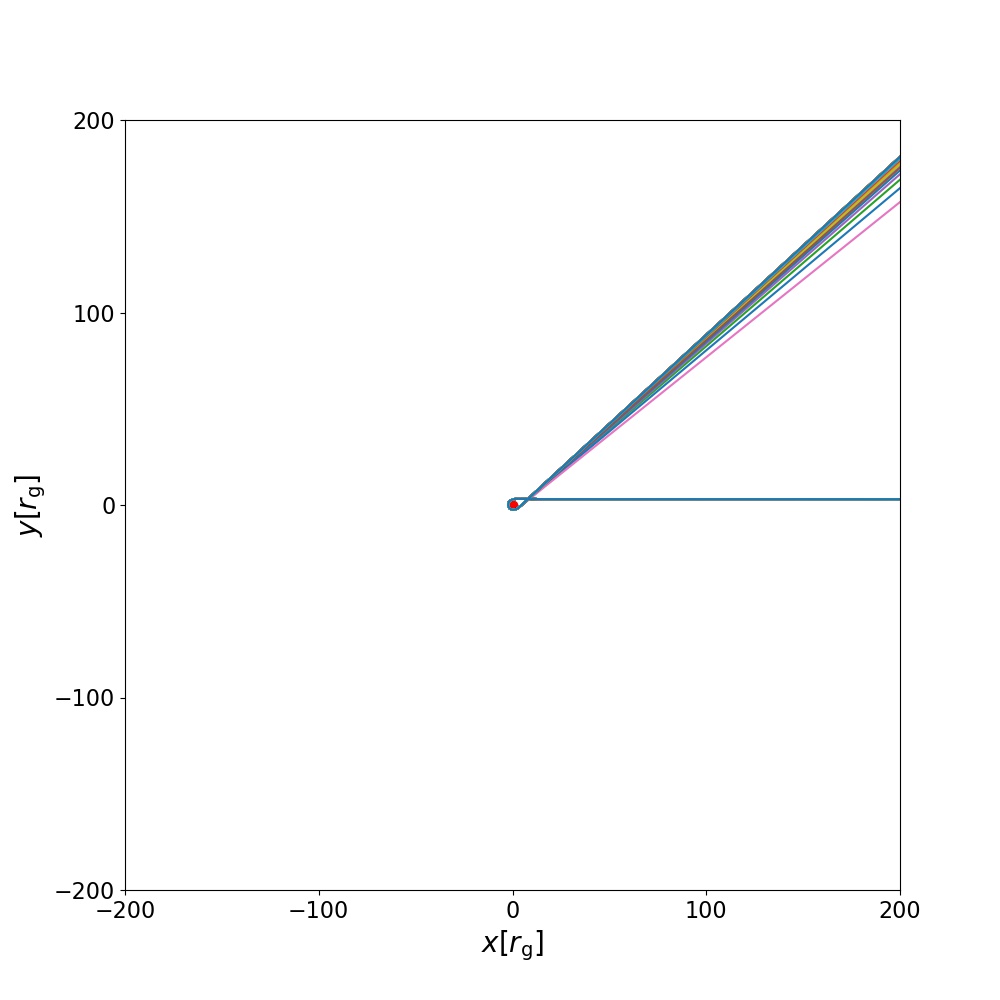}}\\
	\medskip
		\caption{Spatial dispersion induced in the rays propagating through a plasma on a Kerr background geometry. The ray bundle is composed of rays with frequencies linearly spaced between 0.18 and 8 GHz. The BH spin parameter is a maximal $a=0.998$ and $n_0 = 3.5 \times 10^7$ cm$^{-3}$. We set impact parameter $\beta = 0$ such that the rays propagate in the equatorial plane.} \label{fig:example_dispersion}
\end{figure*}

In a vacuum all photons follow the same spacetime geodesic, irrespective of their energy, by virtue of the equivalence principle. However, the presence of plasma means that the spatial trajectory of the ray is now frequency-dependent. Propagating perpendicular to the observer's image plane, initially the rays all take the same spatial path. However, after being deflected by the black hole, different energy photons follow different trajectories, with lower energy photons being more significantly dispersed (compared to the vacuum case) as the plasma frequency term $\omega_{\rm p}$ becomes a greater fraction of the photon frequency (see e.g. Fig. \ref{fig:example_dispersion}). Correspondingly, with an increasing photon frequency the significance of the plasma influence  on the photon trajectories is reduced and the ray path approaches that of the vacuum in the high-frequency limit. Those trajectories which have smaller impact parameters pass closer to the black hole and so traverse a more strongly curved spacetime and also probe increasingly dense  regions of the plasma; consequently the magnitude of spatial dispersion is greater. 

\begin{figure}
	\includegraphics[width=\linewidth]{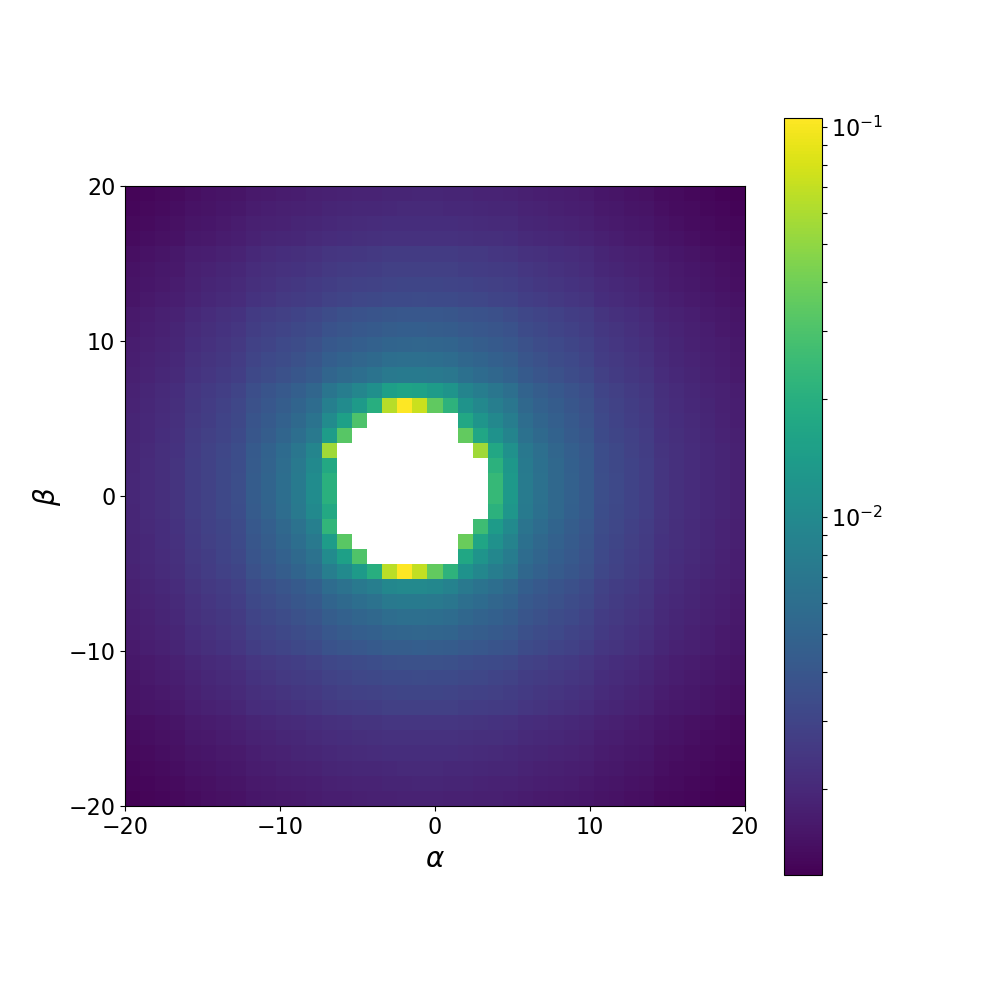}
	\caption{Spatial dispersion $ds$ of a $8.2$ GHz ray relative to the vacuum case as evaluated at $r=1000$ in the outgoing region. The intensity of dispersion increases as rays pass closer to the central BH and probe increasingly dense plasma regions. The white region at the centre denotes impact parameters for which the ray falls below the BH event horizon.} \label{Fig:DispersionMap}
\end{figure}

\begin{figure}
	\includegraphics[width=\linewidth]{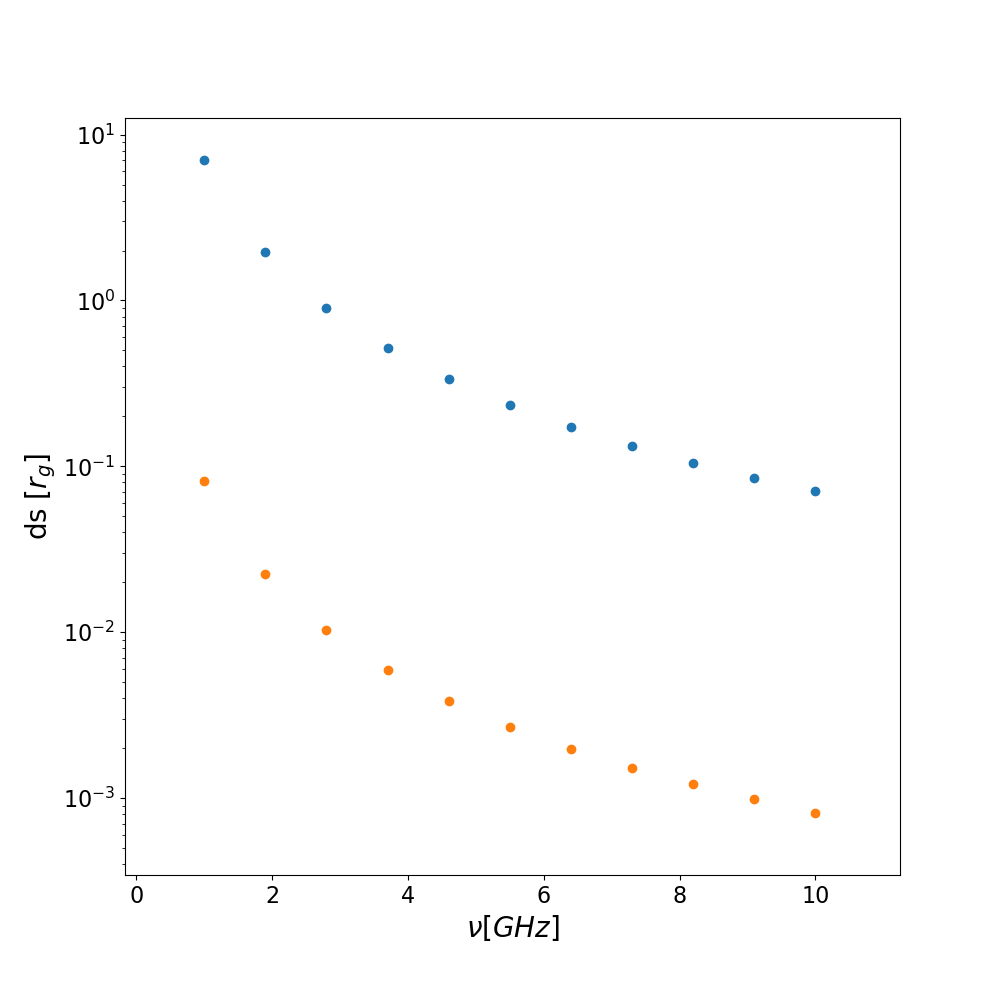}
	\caption{Maximum (blue) and minimum (orange) spatial dispersion $ds$ relative to the vacuum ray at frequencies from 1 to 10 GHz. Spatial dispersion persists even at high radio frequencies, the minimum $ds$ at $\nu = 10$ GHz is $\sim 10^{-3} \, r_{\rm g}$ within the considered parameter space.} \label{fig:cross_frequency}
\end{figure}

\noindent More quantitatively, the degree of spatial dispersion can be described by calculating the spatial difference between the location of the vacuum ray ($x^i_{\rm vac}$) and the location of a ray of particular frequency travelling through a plasma ($x^i_{\rm p} (\nu)$), as evaluated in the outgoing region after being deflected by the black hole, at $r=1000 \, r_{\rm g}$, i.e. 
\begin{eqnarray}
	dx^i =  x^i_{\rm p} (\nu) - x^i_{\rm vac} \ ,
\end{eqnarray}
and we approximate the spacetime on these small scales as being flat such that the metric is Minkowskian, $\eta_{\mu \nu}$. The magnitude quantifying the degree of dispersion, i.e. the deviation from the vacuum path is then the usual expression,
\begin{eqnarray}
	ds^2 = \eta_{i j} dx^i dx^j \ .
\end{eqnarray}
There is considerable uncertainty in the spin parameter of the Galactic centre black hole depending upon the observational approach used to estimate the spin. Observations of quasi-periodic oscillations in the radio emission of Sgr A$^*$ suggest $a \sim 0.44 - 0.65$ \citep{Kato2010a, Dokuchaev2014}, whilst $a \sim 0.996$ based on X-ray lightcurves \citep{Aschenbach2010}. Going forward we take $a=0.6$ as our fiducial value. Clearly, the severity of dispersion is going to be a function of the ray frequency, with the ray path approaching the path in vacuum in the high frequency limit. We examine typical radio frequencies in the GHz regime and inspect the impact parameter space  $|\alpha, \beta | \le 20 \, r_{\rm g}$. The dispersion $ds$ relative to the vacuum is shown in Fig. \ref{Fig:DispersionMap} for a ray with frequency $8.2$ GHz. We can observe a clear trend on the severity of dispersion with impact parameter; rays which pass closer to the black hole suffer greater dispersion since they traverse a more strongly curved spacetime and increasingly dense plasma regions. For a ray at 8.2 GHz, the maximum $ds$ is $\sim 0.11 \, r_{\rm g}$ and the minimum at the edge of the parameter space is $\sim 0.0012 \, r_{\rm g}$. The maximum and minimum $ds$ at frequencies from 1 to 10 GHz is shown in Fig. \ref{fig:cross_frequency}. The dependence of the dispersion on the frequency is visible, with higher frequency rays subject to less dispersion than lower frequency rays. However, even at 10 GHz, the minimum dispersion $ds$ for rays within the impact parameter space is $0.0008 \, r_{\rm g}$. \newline 
\noindent In addition to the dispersion relative to the vacuum path, it is also of interest to quantify the degree of dispersion within a particular observational radio frequency band. We consider both the $C$-band ($4-8$ GHz) and the $X$-band ($8-12$ GHz) and determine the degree of dispersion between the upper and lower elements of the band. For the $C$-band the minimum and maximum dispersion within the prescribed $(\alpha, \beta)$ parameter space is $ (ds_{\rm min}, ds_{\rm max}) = (0.0038, 0.33) r_{\rm g}$ respectively, whilst at the higher frequencies of the $X$-band the intra-band dispersion is reduced: $ (ds_{\rm min}, ds_{\rm max}) = (0.0007, 0.06) r_{\rm g}$. The trend with impact parameter is naturally the same as that displayed in Fig. \ref{Fig:DispersionMap}. Whilst the extent of spatial dispersion does reduce at higher frequencies, we can see that even at the higher frequencies of the $X$-band, the spatial dispersion persists. We are working in a lengthscale of $r_{\rm g}$. Converting to S.I. units, for a BH of Sgr A$^*$ mass, $ds_{\rm min}$ in the $X$-band corresponds to $\sim 4.4 \times 10^6$ meters.  
\subsection{Temporal Dispersion}
In addition to a spatial dispersion, the plasma also induces a dispersion in time. This time delay has two contributing factors. First is the time dispersion of rays of different frequencies which travel along the same path. In a plasma, higher frequency photons will cover the same path in a shorter time than lower frequency photons. Additionally, spatial dispersion means that in fact rays do not follow the same trajectory, but instead follow different spatial paths. Consequently, photons of different frequency have different path lengths, traverse disparate spacetime curvatures and encounter separate regions of the plasma; all of these factors influence the degree of temporal dispersion. \newline 

\begin{figure}
	\includegraphics[width=\linewidth]{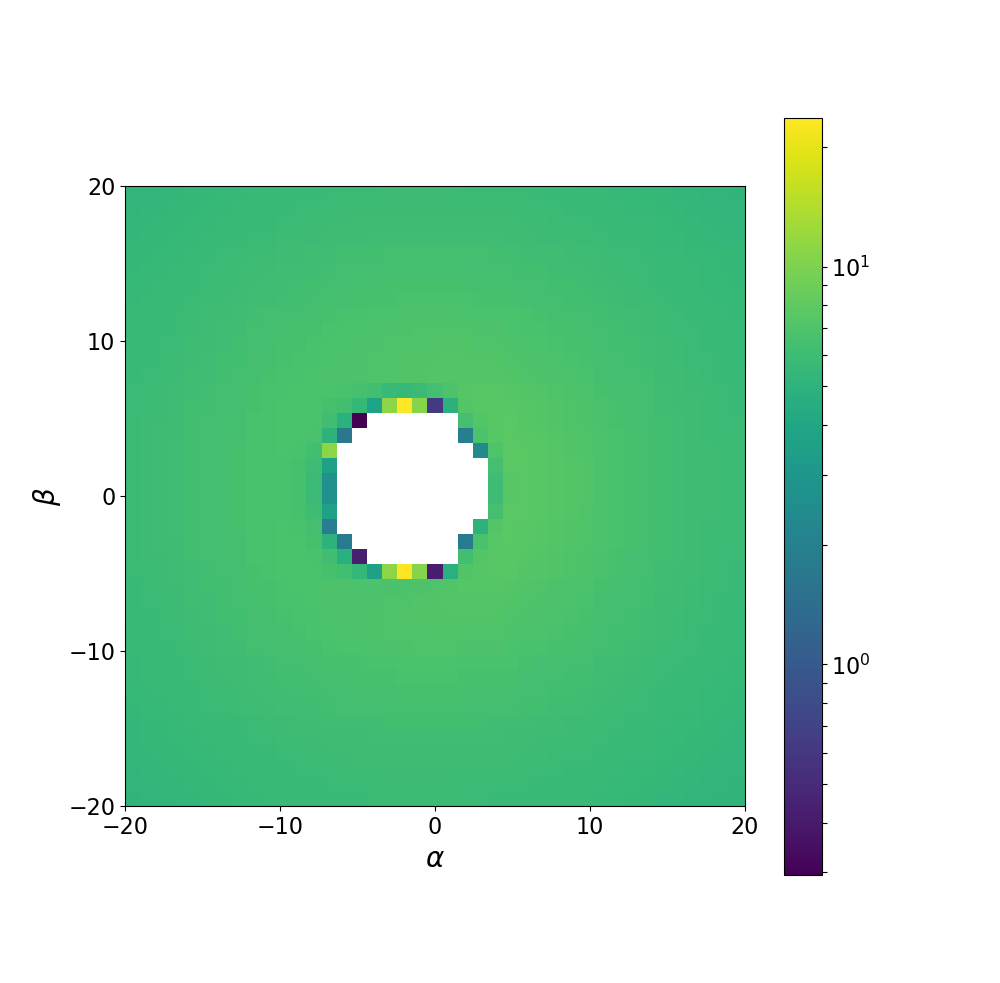}
	\caption{Temporal dispersion $dt$ in milliseconds across the $C$-band radio frequency - i.e. dispersion of an $4$ GHz ray relative to a $8$ GHz ray - as evaluated at $r=1000$ in the outgoing region. The dependence on the impact parameter is much less strong than in the case of spatial dispersion, with typical temporal dispersions of the order $5 \times 10^{-3}$ s. The spin parameter $a = 0.6$ and $n_0 = 3.5 \times 10^7$ cm$^{-3}$.} \label{fig:timemap}
\end{figure}

\noindent We can quantify the extent of temporal dispersion across a frequency band by determining the time delay between the coordinate time of the ray ($x^0$) relative to the ray at the upper frequency of the band ($x_{\rm up}$), i.e.
\begin{eqnarray}
	dt = x^0 - x^0_{\rm up} \ .
\end{eqnarray}
Similar to the preceding analysis regarding spatial dispersion, we evaluate $dt$ in the outgoing region at $r=1000$ and consider the parameter space $|\alpha, \beta | \le 20 \, r_{\rm g}$. For the $C$-band the minimum and maximum temporal dispersion is $ (dt_{\rm min}, dt_{\rm max}) = (0.30, 24)$ ms respectively. The intra-band dispersion is reduced for the higher frequencies of the $X$-band: $ (dt_{\rm min}, dt_{\rm max}) = (0.055, 4.4)$ ms. Similar to the spatial dispersive case, the magnitude of temporal dispersion reduces at higher frequencies and paths further from the central BH. However, the severity of the temporal dispersion is not such a strong function of the impact parameter (Fig. \ref{fig:timemap}); within the considered parameter space the typical $dt \sim 5 $ ms. Again, the dispersion persists even at the higher frequencies of the $X$-band.

\section{Discussion}
\label{sec:discussion}
The detection of a radio pulsar in a compact orbit around a massive black hole has been dubbed the `holy grail of astrophysics' \citep{Faucher2011} for the potential to probe GR in the non-linear, strong-field regime \citep[see e.g.][]{Kramer2004, Liu2012, Psaltis2016}. In addition to detecting such a pulsar-BH system, numerous authors \citep{Wang2009a, Wang2009b, Stovall2011, Nampalliwar2013, Estes2017} emphasize the potential of detecting a pulsar signal for which the path has been strongly bent due to the curvature of the spacetime. Such a beam would propagate directly through the gravitational strong-field regime and so offer a direct probe of this parameter space. Furthermore, the presence of a distribution of masses (e.g. stars) introduces a Newtonian perturbation to the pulsar orbit which may hamper any attempts to use pulsars as a GR apparatus \citep{Merritt2010}. Consequently, it is desirable to time and detect a pulsar as close as possible to the BH, where these external perturbation are negligible, i.e. either at periapsis of some eccentric orbit, or timing some very compact orbit \citep{Liu2012, Psaltis2016}. If a pulsar is observed in these strong-field regions, it is naturally more likely that the pulsar beam may suffer strong bending. For a pulsar in a Keplerian orbit around the BH at the Galactic centre with orbital period $P = 0.1$ years \citep[the typical orbital period below which external perturbations are negligible,][]{Liu2012} and eccentricity $e = 0.9$, then at periapsis the pulsar is only $\sim 80 \, r_{\rm g}$ from the BH event horizon. It therefore seems reasonable that for particular orbital parameters the observed pulsar beam could be strongly bent, although this would naturally only occur for the pulsar on the `far-side' of the black hole and for certain orbital configurations \citep[c.f. inclination of orbit, orientation of the spin and radiation axes etc. See][for a full discussion]{Stovall2011}. \newline 

\begin{figure*} 
	\subfloat[\label{fig:raysa}]{\includegraphics[width=0.45\textwidth]{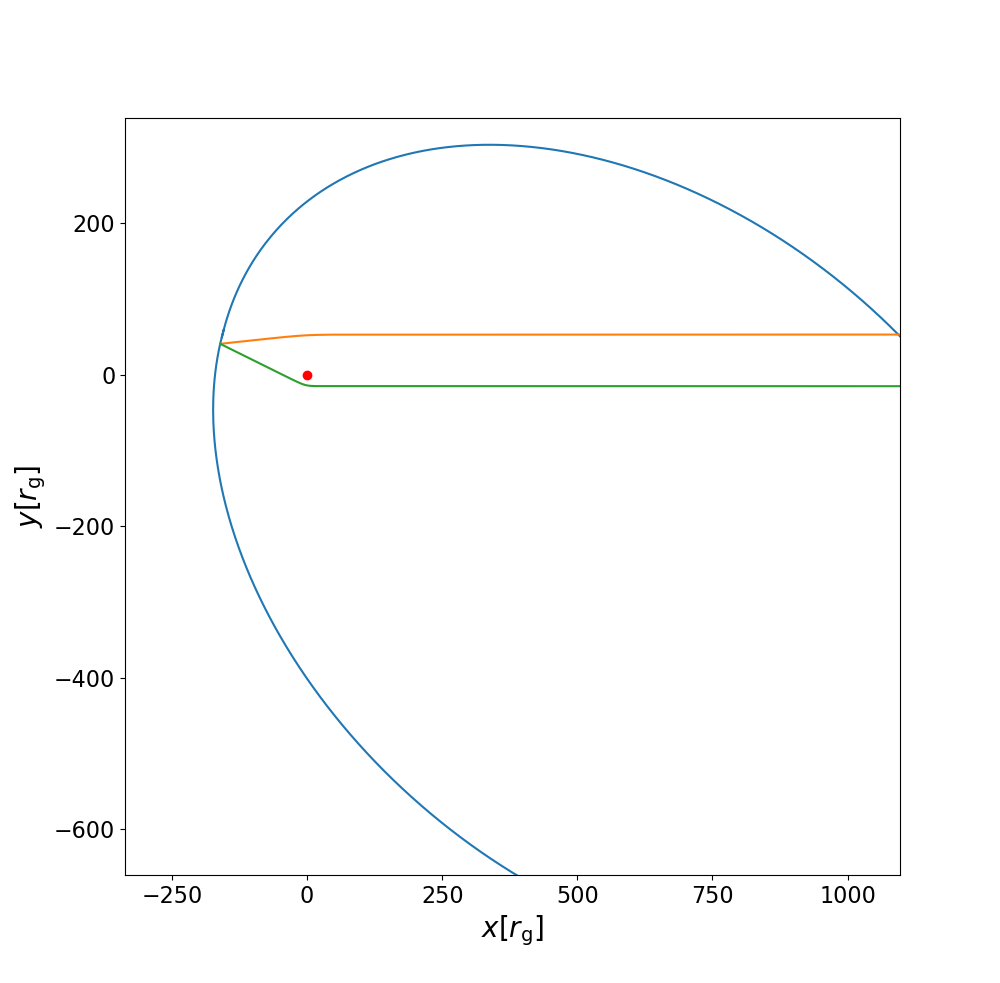}}
	\subfloat[\label{fig:raysb}]{\includegraphics[width=0.45\textwidth]{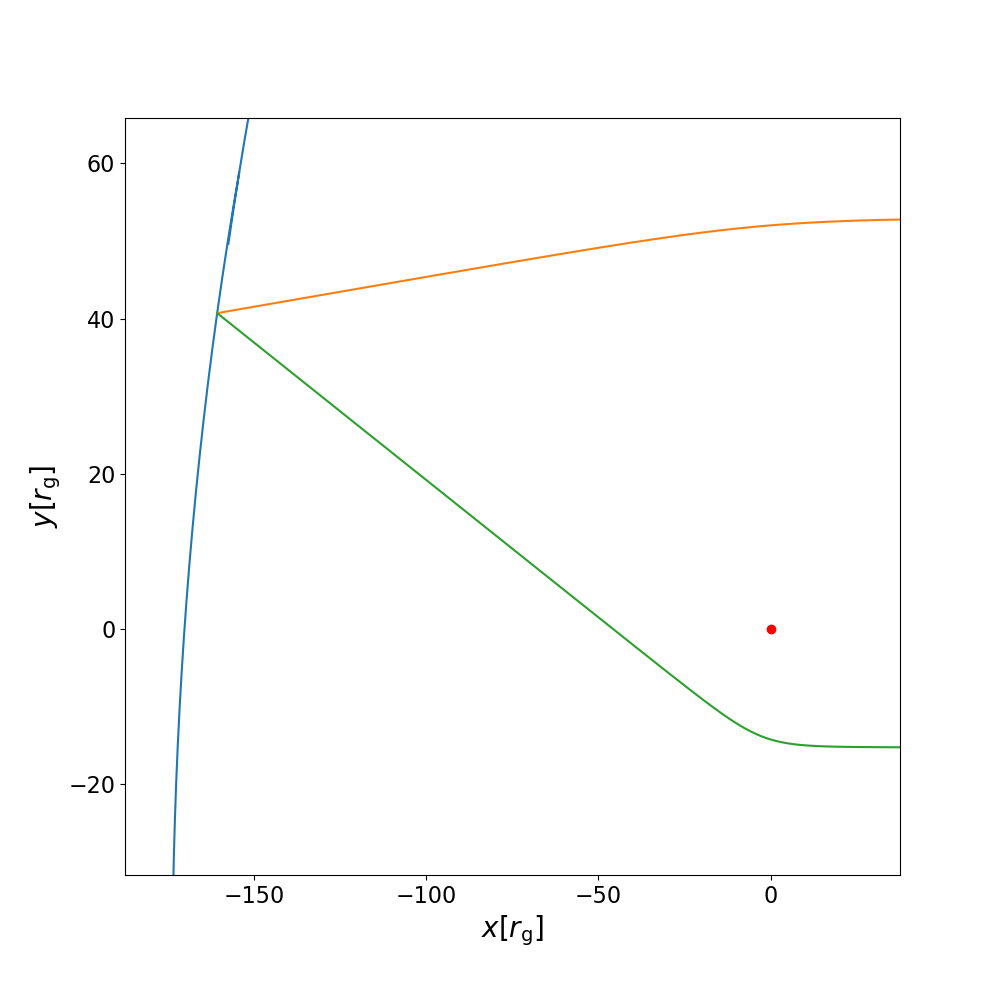}} \\
	\subfloat[\label{fig:raysc}]{\includegraphics[width=0.45\textwidth]{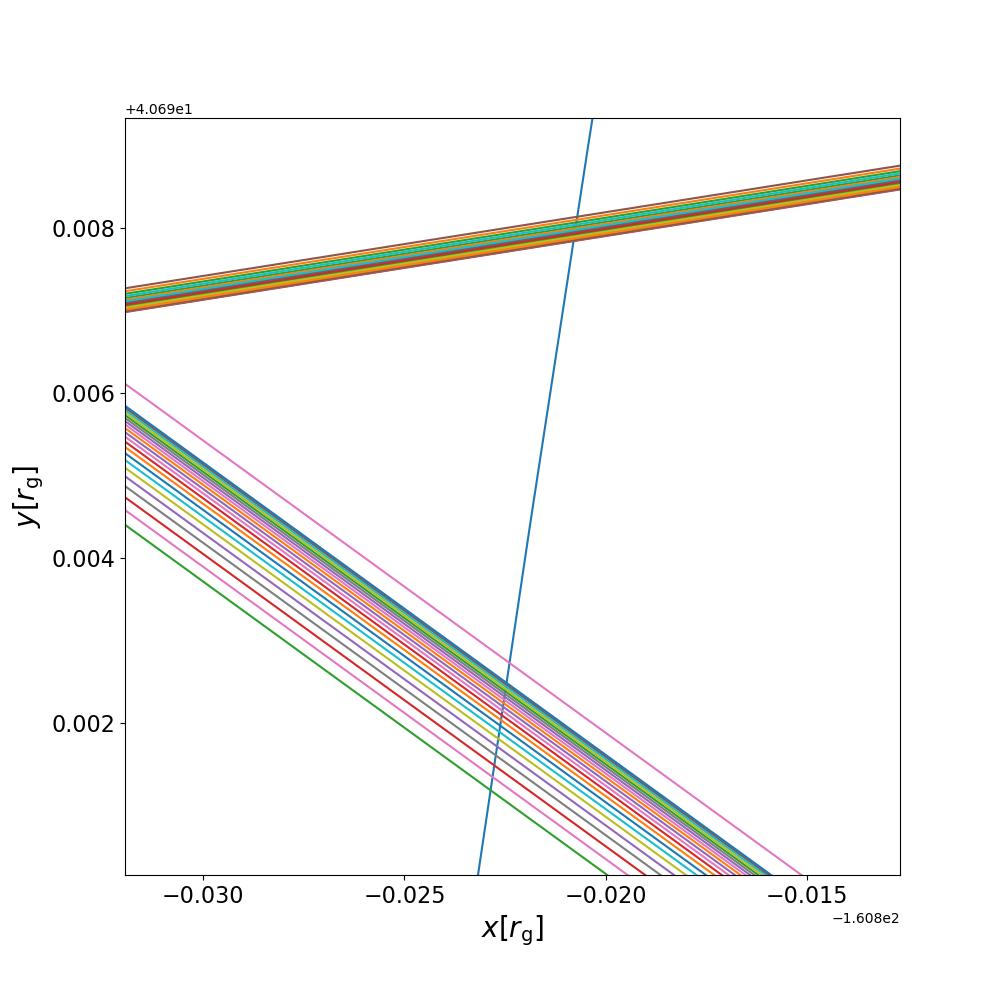}}\\
	\medskip
	\caption{(a) and (b) show the primary and secondary deflected rays in vacuum from a pulsar in a $e=0.8$, $P=0.1$ year orbit, emitted on the far side of the BH. (c) illustrates the spatial dispersion if we integrate a ray bundle in the $C$-band with the same impact parameters as in the vacuum case. The spatial dispersion of these rays in conjunction with the temporal dispersion and the pulsar travel time may further complicate the detection of highly deflected pulsar beams.} \label{fig:rays}
\end{figure*}

\noindent Typical pulsar detection works by attempting to de-disperse the signal due to temporal dispersion induced by the interstellar medium, and then searching for periodicities via a Fourier transform \citep[see][for details of pulsar astronomy]{Lorimer2004}. The degree of temporal dispersion is quantified by the dispersion measure - the integrated electron number density along the ray path. The signal is then folded on the timescale of the periodicity to create a single pulse profile that shows above the noise. If a pulsar ray is deflected in the presence of plasma, this work shows that the ray will exhibit a spatial dispersion. This spatial dispersion is naturally coupled with a temporal dispersion and so has a series of implications for the observability of such deflected beams. Firstly, we have presented this work using a `backwards-in-time' approach where the ray is integrated from the observer to the pulsar. If we instead shift to a `forwards-in-time' interpretation then evidently if the spatial dispersion is sufficiently large it could cause the pulsar signal to not be visible in certain frequency bins since the ray path is curved such that it never reaches the observer. Furthermore, those rays which do reach the observer have each traversed a different spatial path and so are subject to different time delay due to differences in the ray path and also different dispersion measures along said path. Returning to the `backwards-in-time' outlook, to successfully perform a search in Fourier space requires that the received pulse period is constant. However, due to spatial dispersion the total received signal across the observation frequency band is the convolution of different energy rays emitted at differed orbital phases, subject to gravitational, relativistic and line of sight effects which may introduce additional difficulties in detecting deflected beams from compact pulsar systems (see Fig. \ref{fig:rays}). Traditionally, systems for which the observed pulse period is changing over the observation time (i.e. highly accelerated systems, c.f. binary pulsars) are detected via so called `acceleration searches' \citep[see e.g.][]{Dimoudi2015}. However, such an approach would not work for the detection of bent pulsar rays subject to spatial dispersion and an alternative methods may be necessary.

\begin{figure}
	\includegraphics[width=\linewidth]{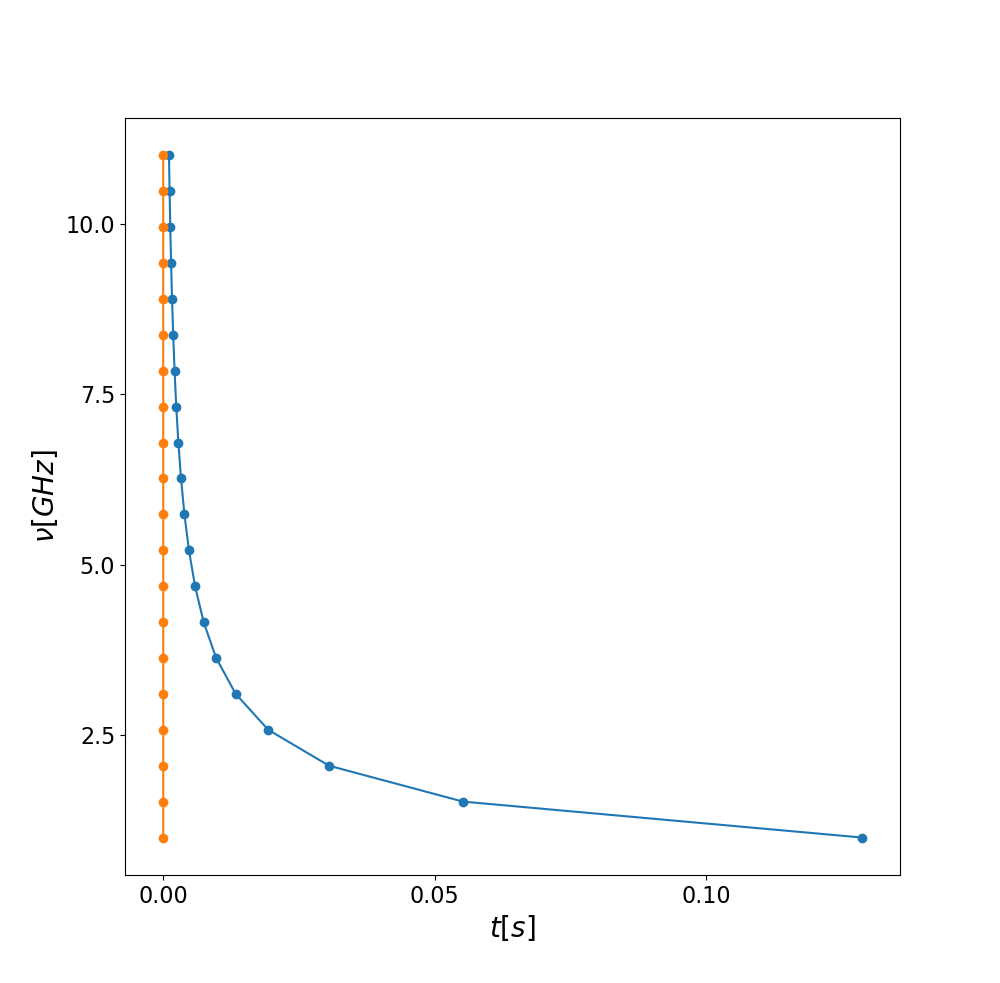}
	\caption{Time-frequency profile in vacuum (orange) and subject to plasma effects (blue). The profile that suffers from plasma effects is retarded with respect to the vacuum case. Moreover, the combination of spatial and temporal dispersion means that low frequency photons are delayed with respect to higher frequency ones, a phenomenon not observed in the vacuum case. We use the electron density profile $n = n_0 r^{-1.1}$, with $n_0 = 3.5 \times 10^7$ cm$^{-3}$. The BH mass is $4.3 \times 10^6 M_{\odot}$and spin parameter $a=0.6$.} \label{fig:profileeg}
\end{figure}

The strongly deflected beams typically belong to the class of `secondary rays' that pass close to the black hole. There also exists a `primary ray' which is subject to less or negligible gravitational bending. However, for certain orbital configurations, both the secondary and primary rays can exhibit gravitational bending. This is shown in Fig \ref{fig:raysa} for emission from a pulsar on the far-side of the BH, close to periapsis. Consequently both the primary and secondary rays can be subject to the resulting spatial and temporal dispersions induced by the plasma. We consider an orbit with Keplerian period $P=0.1$ years, eccentricity $e=0.8$, and inclination with respect to the black hole spin axis $\Theta = 15^{\circ}$. In the vacuum case we can see there exist two rays received by the observer from the pulsar at this location (Figs \ref{fig:raysa}, \ref{fig:raysb}). Each of these rays follows a different trajectory before being detected by the observer, and the secondary ray suffers much greater gravitational bending than the primary ray. The secondary ray is retarded in time with respect to the primary ray due to the increased spacetime path. Each of these vacuum rays has a particular set of impact parameters $(\alpha, \beta)$. We integrate a ray bundle in the $C$-band through a plasma in a Kerr spacetime with the same impact parameters of the primary/secondary rays in the vacuum case (Fig \ref{fig:raysc}). A clear spatial dispersion is displayed in both the primary and secondary rays at the considered section of the pulsar orbit. The spatial dispersion is more severe in the primary ray than in the secondary ray, but is evident in both on the scale of $\sim$ milli-$r_{\rm g}$. Consequently, the received deflected signal from a pulsar in these regions is the sum of emission at different orbital phases, subject to disparate relativistic and gravitational shifts, temporal dispersions and path lengths. Therefore, for the purposes of detecting deflected beams in this region, spatial dispersion introduces additional complications that may need to be accounted for if such a search is to be successful. Figure \ref{fig:profileeg} illustrates the difference between the time-frequency profile in vacuum and when subject to plasma effects.

To summarise, evidently spatial dispersion only plays a role if the pulsar is on the far side of the BH and the ray passes sufficiently close to the event horizon. It will therefore not be important for wide orbits, emission on the near side of the BH or emission far from periapsis and is highly dependent on the pulsar orbital configuration. However, any discussion regarding the detection of gravitationally deflected pulsar beams \citep[e.g.][]{Wang2009a, Wang2009b, Stovall2011, Nampalliwar2013,Estes2017} must take into account spatial dispersion since it introduces additional complexities in the detection of such bent rays from pulsars. The spatial dispersion can occur for both primary and secondary rays - subject to caveats regarding the orbital configuration. The magnitude of dispersion is reduced as $\omega_{\rm p}$ becomes negligible. Consequently observations at higher frequencies are desirable and would complement the existing approach to mitigate the known problems of scattering \citep[see e.g.][]{Macquart2010,Spitler2014,Bower2015,Rajwade2017}. However, even in the $X$-band frequencies the dispersion persists non-negligibly and pulsars typically have steep radio spectra which makes detection at higher frequencies more problematic. Consequently, observations with the increased sensitivities of the next generation of radio telescopes (e.g. FAST, SKA) may be required. At lower frequencies, scattering will dominate over the dispersive plasma effects for 'near-side' rays. Conversely,  for 'far-side' gravitationally bents rays the dispersive effects would likely outweigh scattering effects, which typically have magnitudes on the order of $\mu $s \citep[see e.g.][]{Palliyaguru2015}.

\section{Acknowledgments}
We thank the anonymous reviewer for their helpful comments. We thank Denis Gonzalez-Caniulef for useful remarks and discussion. TK wishes to thank Marta Burgay, Andrea Possenti and Alessandro Corongiu for helpful discussion on the details of pulsar observation. TK also acknowledges financial support form COST Action: CA16214, funding a visit to INAF Cagliari. 

\bibliographystyle{mnras}
\bibliography{paper2}

\appendix
\section{Initial conditions}
To determine the initial conditions of rays starting on the observer's grid, we follow the methods described in \cite{YounsiThesis,Pu2016} by transforming from the observer coordinate system $\mathbf{x'}$ to the BH coordinate system $\mathbf{x}$, as,
\begin{enumerate}
	\item Rotate clockwise by $(\pi - \theta_{\rm obs})$ about the $x'$-axis ($R_{x'}$)
	\item Rotate clockwise by $(2\pi - \phi_{\rm obs})$ about the $z'$-axis ($R_{z'}$).
	\item Reflect in the plane $y'=x'$ ($A_{y'=x'}$).
	\item Translate $\mathbf{\bar{x'}}$ so that the origins of both coordinate systems coincide ($T_{x'\rightarrow x}$)
\end{enumerate}
The net transformation is then
\begin{eqnarray}
	\mathbf{x} &= A_{y'=x'}R_{z'}R_{x'} \mathbf{x'} +T_{x'\rightarrow x} \\ 
	&= \left( \begin{array}{c}
		\mathcal{D}(y',z') \cos \phi_{\rm obs} -x' \sin\phi_{\rm obs}  \\
		\mathcal{D}(y',z') \sin \phi_{\rm obs} +x' \cos\phi_{\rm obs}  \\
		(r_{obs}-z') \cos \theta_{\rm obs} + y' \sin \theta_{\rm obs}  \end{array} \right)  \ ,
\end{eqnarray}
where $\mathcal{D} = (\sqrt{r_{\rm obs}^2 +a^2} -z') \sin \theta_{\rm obs} - y' \, \cos \theta_{\rm obs}$. Then transform from Cartesian to Boyer-Lindquist coordinates, 
\begin{eqnarray}
	r = \frac{\sqrt{w+\sqrt{w^2+4a^2z^2}}}{2 } \ ;  \\ 
	\theta = \arccos \left( \frac{z}{r}\right) \ ;   \\ 
	\phi = \arctan2(y,x)  \ , 
\end{eqnarray}
where $w = x^2 +y^2 +z^2 -a^2$. 
This defines the initial $(r,\theta,\phi)$ for a photon on the observer grid. 

\noindent We then determine the initial velocities of the ray. Since each ray arrives perpendicular to the image plane, 
$(\dot{x'}, \dot{y'}, \dot{z'}) = (0,0,1)$. 
Consequently, the velocity components in the black hole frame are given by 
\begin{eqnarray}
	\dot{x} &= \left( \begin{array}{c}
		-\sin \theta_{\rm obs} \cos \phi_{\rm obs}  \\
		-\sin \theta_{\rm obs} \sin \phi_{\rm obs}  \\
		-\cos \theta_{\rm obs}  \end{array} \right)   \ . 
\end{eqnarray}
Converting to Boyer-Lindquist coordinates gives expressions for $(\dot{r}, \dot{\theta}, \dot{\phi})$ 
in the black hole frame:
\begin{eqnarray}
	\dot{r} = - \frac{-r\mathcal{R} \sin \theta \sin \theta_{\rm obs} \cos{\Phi} 
		+ \mathcal{R}^2\cos \theta \cos \theta_{\rm obs}}{\Sigma}  \\ 
	\dot{\theta} = \frac{r \sin \theta \cos \theta_{\rm obs} 
		- \mathcal{R}\cos \theta \sin \theta_{\rm obs} \cos \Phi}{\Sigma}   \\ 
	\dot{\phi} = \frac{\sin \theta_{\rm obs} \sin \Phi}{\mathcal{R} \sin \theta}
\end{eqnarray}   
where $\mathcal{R} = \sqrt{r^2 +a^2}$ and $\Phi = \phi - \phi_{\rm obs}$. 
This completely defines our initial conditions.

% Don't change these lines
\bsp	% typesetting comment
\label{lastpage}
\end{document}